# Particle Identification in the NIMROD-ISiS Detector Array

S. Wuenschel, K. Hagel, L.W. May, R. Wada, S.J. Yennello

*Texas A&M University Cyclotron Institute College Station TX 77843*

**Abstract.** Interest in the influence of the neutron-to-proton (N/Z) ratio on multifragmenting nuclei has demanded an improvement in the capabilities of multi-detector arrays as well as the companion analysis methods. The particle identification method used in the NIMROD-ISiS $4\pi$ array is described. Performance of the detectors and the analysis method are presented for the reaction of $^{86}$Kr+$^{64}$Ni at 35MeV/u.

**Keywords:** Isotopic Resolution, Particle Identification.
**PACS:** 29.85.-c; 29.85.Fj

## INTRODUCTION

The first generation of multi-detectors focused on complete collection of charged particles produced in an interaction. In general, these detectors focused on elemental resolution and provided little isotopic information about the event. An impressive body of studies came from these detectors. Recently, interest has evolved to include understanding the isotopic degree of freedom. Thus data analysis methods on large-scale detector arrays must evolve as well.

### Improving Isotopic Resolution

Isotopic resolution can be improved in two ways: improve detector resolution or improve data analysis methods. Figure 1 depicts a super telescope from the NIMROD-ISiS array. Particles resulting from a reaction enter from the left. The first detector that they pass through is a 150μm thick Si detector. If the particle has sufficient energy it will then pass into, and perhaps through, a 500μm Si detector and on into the CsI crystal. The CsI crystal is read using a photomultiplier tube. Note that some telescopes do not incorporate the second silicon detector. Thus, there are three possible sources of particle identification in a super telescope such as Figure 1. Z=1,2 particles may be identified by pulse shape discrimination on the CsI light output (see Figure 2 top). Heavier particles are isotopically identified in the Si-CsI or Si-Si (Figure 2 middle and bottom respectively) by the ΔE-E method. The Si-Si plot (Figure 2 bottom) exhibits a band on the left resulting from particles punching through both Si detectors. Data for the analysis is from the reaction of $^{86}$Kr+$^{64}$Ni at 35MeV/u.

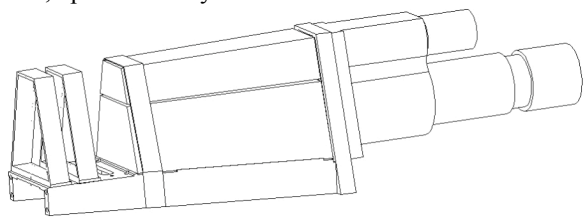

**FIGURE 1.** Super telescope module of the NIMROD-ISiS array. This module incorporates two (150μm and 500μm respectively) Si wafers in front of the CsI crystal. The CsI is read out by a photomultiplier tube. Fragments enter from the left.

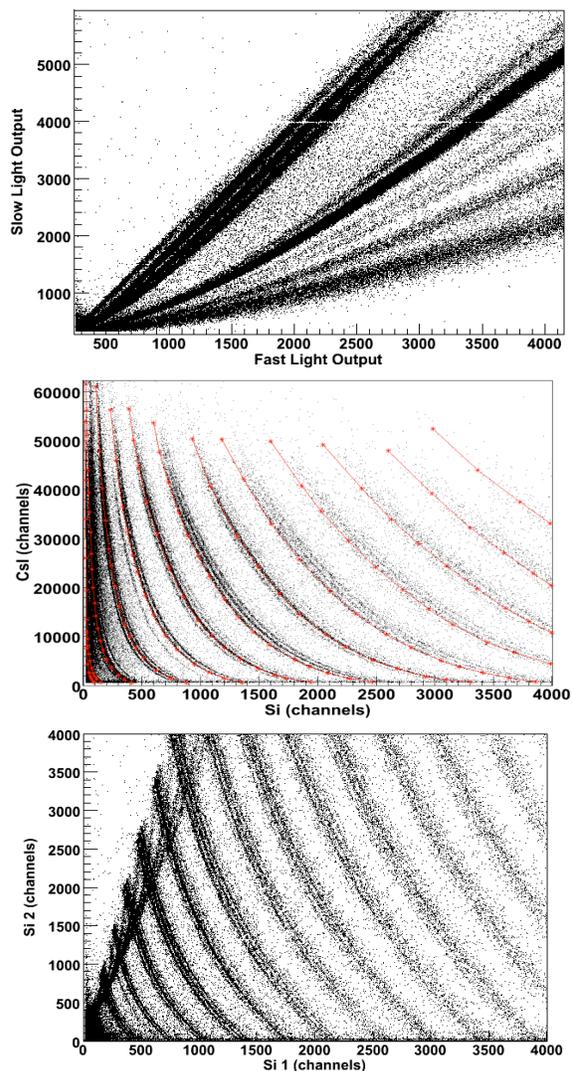

**FIGURE 2.** Sources of particle identification in a detector module (Figure 1). Top) CsI pulse shape discrimination on the light output of the CsI crystal Middle) Energy deposited in the 500μm Si vs the energy deposited in the CsI. Bottom) Energy deposited in the 150μm Si vs 500μm Si (Color online)

Data analysis for large detector arrays is very complex. A major consideration during analysis is to choose a method that is reasonable considering the sheer number of detection elements that must be addressed.

## Particle Identification Method

There are two primary methods used for particle identification in large arrays. The first method requires the user to manually choose gates for the visibly resolved data. The second [4] method linearizes the data and allows the user to project the straightened lines into a one-dimensional plot. Linearization utilizes lines carefully chosen to follow the data trends. The data is then straightened using a calculation of the distance between the data point and the chosen lines.

Two methods exist for choosing these lines. These lines may be chosen by hand or generated by an energy loss code. Lines chosen by an energy loss code work well for low Z elements but gradually begin to fail for higher Zs and for the high and low energy regions of data [4].

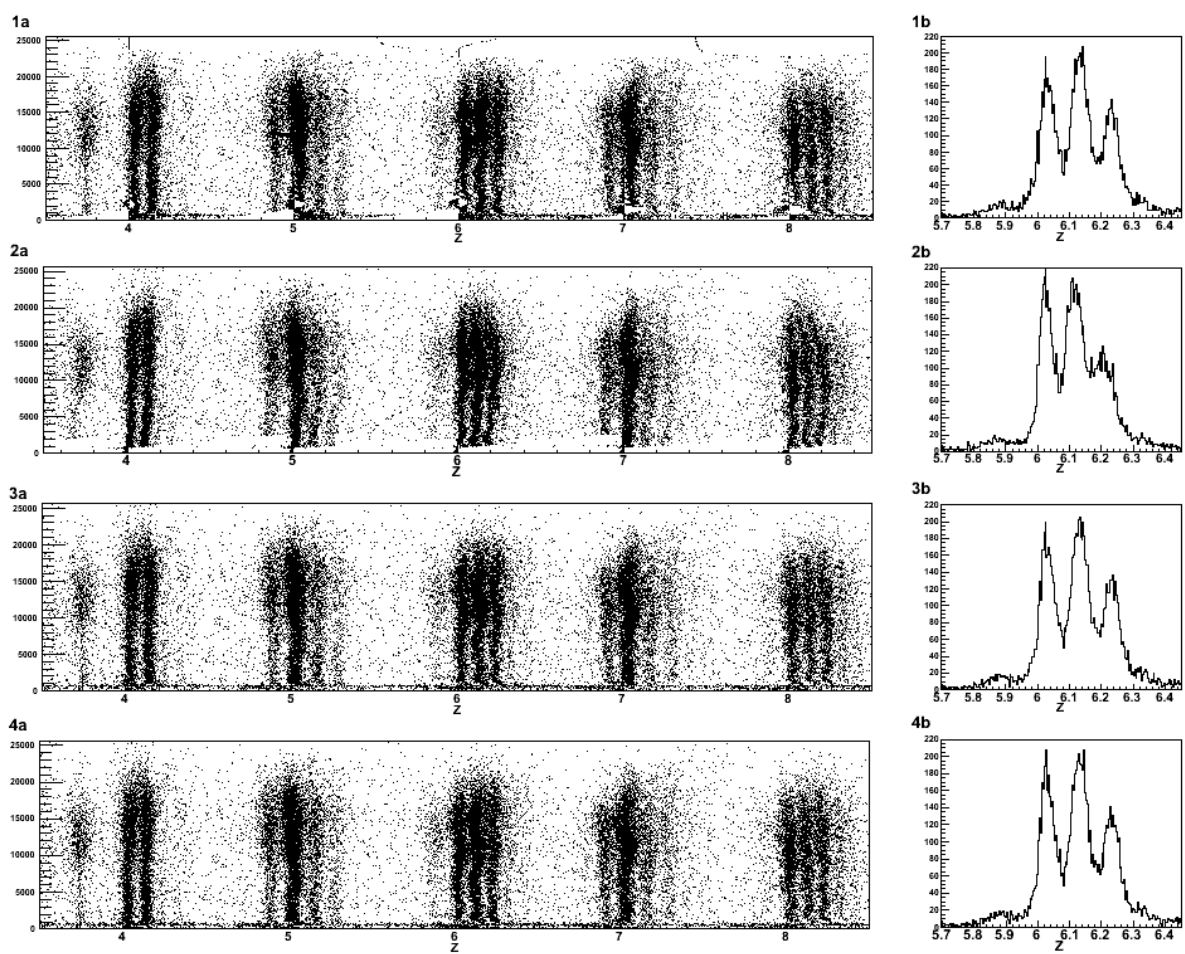

**FIGURE 3**. Left hand panels are linearized plots of CsI channel versus the normalized linearization distance. 1) Horizontal distance. 2) Vertical Distance. 3) 45-degree Distance. 4) Point-to-Curve Distance. Right hand plots depict the projection of the Carbon isotopes for each linearization method.

Lines may be chosen by hand, but this is labor intensive. However, if retaining the maximum amount of isotopically resolved data is the main goal, hand chosen lines are currently the best option.

The data presented in this paper is analyzed using the linearization procedure with user chosen lines. Example lines are shown superimposed on data in the middle panel of Figure 2. In general, one line was constructed for each element. However, to optimize the resolution of Z=1 isotopes, two lines were used.

A graphical user interface was developed to facilitate the users interaction with the data plots. The GUI also allows the user to test the quality of the chosen lines as progress is made. The specifics of the GUI are not important; but rather, the value lies in the ease of interaction such an interface can provide.

*Linearization Methods*

Once lines are chosen and optimized (Figure 2 – middle panel), the data is linearized. This is done by calculating the distance between the data point and the closest line and normalizing with respect to the distance between the two closest lines. The specific method of calculating the distance can have a significant affect on the quality of the linearization. Figure 3 shows the resulting linearization from Si-CsI data using four different distance calculation methods. The y-axis is the channel number of the signal from the CsI light output. The x-axis is the normalized linearization distance resulting from the respective methods. Panels 1a and 2a are based on horizontal and vertical distances to the lines, where the distance to the line is calculated in either the x or y direction. These two distance methods are in practice similar processes with differing results due to their relationship to the Si axis. Resolution in Si detectors is innately better than that in CsI crystals.

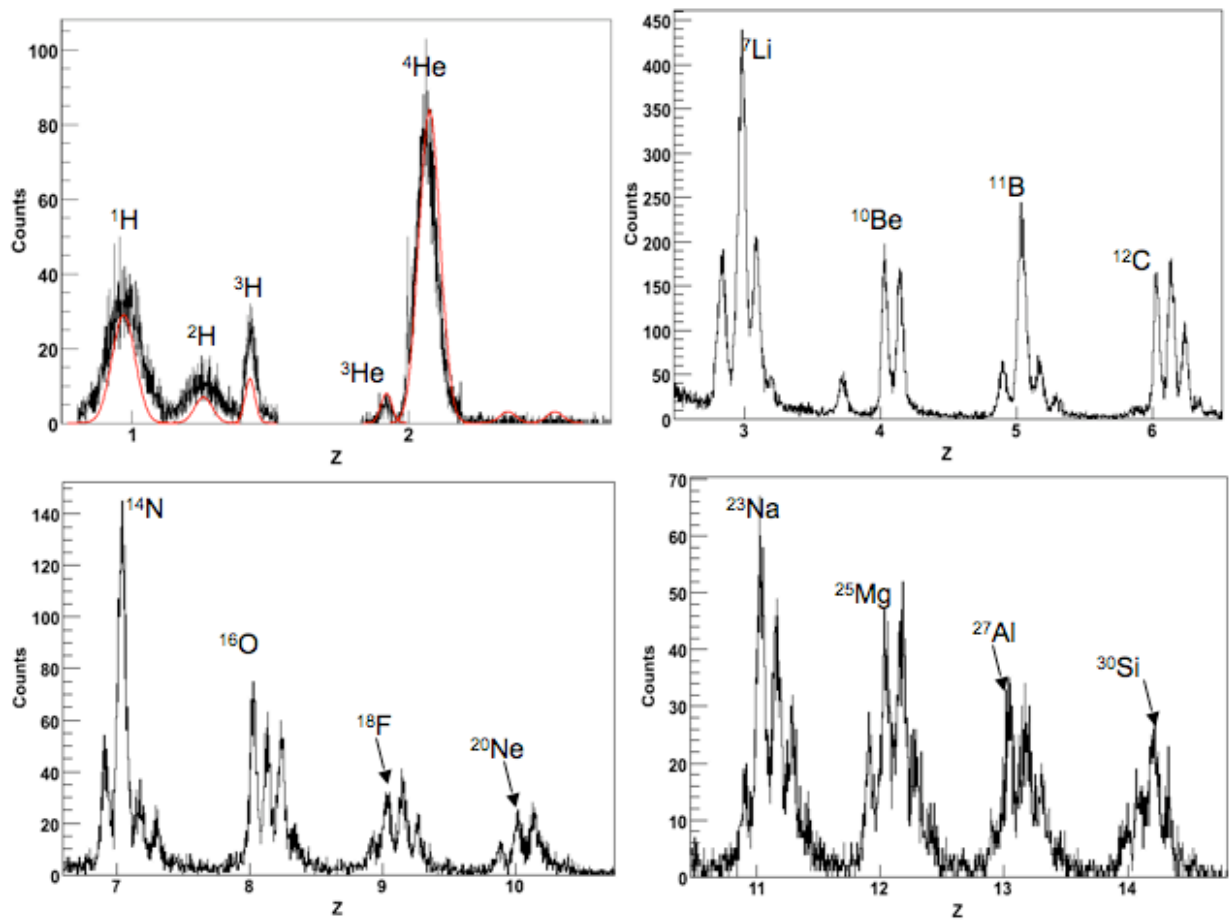

**FIGURE 4.** Results of the linearization procedure. Z=1,2 are fitted with Gaussian function (1) depicted by the red lines (color online) Z=1,2 are identified in the CsI Fast-Slow, Z=3-10 are representative of Si-CsI data, and Z= 11-14 were derived from a Si-Si data plot.

Thus, for plots with similar curvature, calculating distance along the Si axis will generally be more precise than along the CsI axis. The Horizontal distance method (Panel 1a) clearly has difficulties below the CsI channel ~5000. This is the most horizontal portion of the plot. The vertical distance method (Panel 2a), on the other hand, has significant problems above channel ~5000; i.e. in the most vertical portions of the curves. In addition to the problems at higher CsI channel, the vertical distance method imparts an increasing curvature as Z increases. This can be seen in the projection of the carbon isotopes; most notably the [14]C peak in Panel 2b. Panel 3a is the result of distance being calculated along a modified 45 degree line [5]. Panel 4a is an absolute distance calculation between the data points and the nearest point on the chosen lines. The performances of the 45-degree line and the point-to-curve methods are very similar. The point-to-curve method should be better for very flat regions of data; however, as implemented here, it shows no significant improvement. Both the point-to-curve and the 45-degree line methods perform better for extremity data than the horizontal or vertical method.

*Gaussian Probability Function*

After linearization, data is then projected onto a one-dimensional plot producing quasi-Gaussian peaks. Figure 4 shows the intensity of the peak versus the normalized distance to the two closest lines. The elemental groups are widely spread. Within each elemental group, the isotopic resolution is clear through Z=14. The most prominent isotope of each element is labeled for reference. The isotopic peaks within an element were fitted with Gaussian functions that were then correlated to masses. Analysis in this manner provides the user with a means of defining confidence in the quality of the isotopic designation. This information can be used to assess confidence in the N/Z of the reconstructed source.

$$\Pr = \frac{G_i}{\sum\limits_{i}^{i=N_{peaks}} G_i} \qquad (1)$$

(1) Probability Function - G is the value of the chosen Gaussian at the x-axis position

A simple probability (1) was constructed for defining event-by-event particle designation.

One consideration, in this method, is that the user defines the mass designation of each peak. Using code generated lines would have the benefit of the mass being a byproduct of the distance calculation. However, mass designations may be checked against published distributions such as those of Mocko and Souliotis [6,7].

## Performance and Conclusion

The performance of this method can be seen in the composite Figure 4. Note that in this plot Z=1,2 are identified in the CsI Fast-Slow, Z=3-10 are representative of Si-CsI data, and Z= 11-14 were derived from a Si-Si data plot. It is interesting that the [3]H peak is narrower than the other two H peaks. This is an artifact of the linearization using two lines rather than one. It is not a true feature of the data.

The NIMROD-ISiS array provides isotopic identification across a broad range of angles, elements, and energy. The detector quality, coupled with the analysis method described here, result in a maximum amount of isotopic information available during the analysis phase.

## ACKNOWLEDGMENTS

This work was supported by the United Stated Department of Energy under grant DE-FG03-93ER40773 and by the Robert A. Welch Foundation under grant A1266.